\documentclass[onecolumn,secnumarabic,amssymb, amsmath, nofootinbib,tightenlines,
nobibnotes, aps, prl,epsfig]{revtex4}
\usepackage{graphicx}
\usepackage{dcolumn}
\usepackage{bm}
\begin{document}
\preprint{APS/123-QED}
\title{Heavy quark structure functions from unifying the color dipole picture and double asymptotic scaling approaches }

\author{G.R.Boroun}%
 \email{boroun@razi.ac.ir }
\affiliation{ Department of Physics, Razi University, Kermanshah
67149, Iran}
\date{\today}
\begin{abstract}
We present an analysis of the heavy quark structure functions from
the $k_{t}$ factorization scheme, using unifying the color dipole
picture and double asymptotic scaling approaches at small $x$. The
gluon distribution is obtained from the
Golec-Biernat-W$\ddot{\mathrm{u}}$sthoff (GBW)  and Bartels,
Golec-Biernat and Kowalski (BGK )models. The main elements are
based on the color dipole picture (CDP) and  the generalized
double asymptotic scaling (DAS) approach for usual parton
distribution functions (PDFs). The comparisons with the HERA data
are made and predictions for the proposed LHeC and FCC-he
colliders are also provided in a wide range of  the transverse
separation $r$. In particular, the ratio
$R^{h}=F_{L}^{h}/F_{2}^{h}, h=c,b,t$ is well described by the
dipole models and is sensitive to the collider energies from HERA
until FCC-he. We derive correlated bounds on the ratio
$F^{c}_{2}/F_{2}$ and $F^{b}_{2}/F_{2}$ and compared them with the
BGK and IP-sat models. The uncertainties are due to the
renormalization and factorization scales at large and low $r$
values. The Sudakov form factor into the heavy quark structure
functions is incorporated and the results are considered, which
are dependent on the
hard scale in a wide range of the transverse separation $r$.\\

\end{abstract}
 \pacs{***}
\maketitle
\subsection{I. Introduction}

The vector-meson-dominance (VMD) model [1,2] is an old idea that
the scattering of a highly-energetic photon on a hadron may
essentially be considered as a strong interaction process whenever
a photon couples to hadrons it first converts to the vector mesons
with universal coupling constants [3]. The VDM model was applied
to deep inelastic scattering (DIS) on the assumption that the
photon fluctuates into a series of vector mesons which
subsequently scatter off the proton [4]. A similar idea, which is
motivated to a large extent by perturbative quantum chromodynamics
(pQCD), is the color dipole model (CDM) [5], which provides a
successful description of deep inelastic scattering processes in a
wide range of the kinematic variables [6]. The QCD color dipole
formalism provides an intuitive description of inclusive and
exclusive processes in electronproton ($ep$) and lepton-nucleus
($lA$) scattering at high energies [7]. Although our knowledge of
the proton structure at small-$x$ is very limited, novel
opportunities will be opened at new-generation facilities
(Electron-Ion Collider(EIC), High-Luminosity Large Hadron Collider
(HL-LHC), Forward Physics Facility (FPF)). Combining the
information coming from dipole cross sections and
$p_{T}$-unintegrated densities could play an important role. In
particular, polarized amplitudes and cross sections for the
exclusive electroproduction of $\rho$ and $\phi$ mesons at the
Hadron-Electron Ring Accelerator (HERA) and the EIC are very
sensitive to the unintegrated gluon distribution (UGD) model
adopted, whereas forward Drell-Yan dilepton distributions at the
Large Hadron Collider beauty (LHCb) are very sensitive to
next-to-leading logarithmic corrections.\\
In the color dipole picture (CDP) the absorption of a virtual
photon on the proton $\gamma^{*}+p~{\rightarrow}X$, is motivated
by perturbation theory and describes photon-proton scattering as a
two-step process. Firstly, the virtual photon dissociates into a
quark-antiquark pair (a $q\overline{q}$ dipole) and subsequently
the pair interacts with the proton, which is a purely hadronic
reaction [8]. The CDP, at small $x$, gives a clear interpretation
of the high-energy interactions, where is characterized by high
gluon densities because the proton structure is dominated by dense
gluon systems and predicts that the small $x$ gluons in a hadron
wavefunction should form a Color Glass Condensate [9,10].\\
In the high energy, $s{\gg}Q^2{\gg}\Lambda^{2}_{QCD}$, regime
these two processes are factorized, and the total cross section
can be written as [5]
\begin{eqnarray}
\sigma_{L,T}^{\gamma^{*}p}(x,Q^{2})=\int dz d^{2}\mathbf{r}
|\Psi_{L,T}(\mathbf{r},z,Q^{2})|^{2}\sigma_{\mathrm{dip}}({x},\mathbf{r}),
\end{eqnarray}
where  DIS cross section is factorized into a light-cone wave
function and a dipole cross section. Indeed, the scattering
between the virtual photon $\gamma^{*}$ and the proton is seen as
the color dipole where the transverse dipole size $r$ and the
longitudinal momentum fraction $z$ with respect to the photon
momentum are defined. The subscripts $L$ and  $T$ referring to the
transverse and longitudinal polarization state of the exchanged
boson. Here $\Psi_{L,T}$ are the appropriate spin averaged
light-cone wave functions of the photon and
$\sigma_{\mathrm{dip}}({x},r)$ is the dipole cross-section which
related to the imaginary part of the $(q\overline{q})p$ forward
scattering amplitude. The variable $z$, with $0\leq z \leq 1 $,
characterizes the distribution of the momenta between quark and
antiquark. The square of the photon wave function describes the
probability for the occurrence of a $(q\overline{q})$ fluctuation
of transverse size with respect to
the photon polarization.\\
The key feature is the connection of the dipole cross section to
the integrated gluon distribution. The parton saturation models
shed light on the behavior of the gluon density at very low $x$
and this knowledge is crucial for instance to describe the
exclusive processes in $ep$ and $eA$ collisions [7]. The dipole
cross section is related to the unintegrated gluon distribution
[12]
\begin{eqnarray}
\sigma(x,\mathbf{r})=\frac{8\pi^{2}}{N_{c}}\int\frac{dk_{t}}{k_{t}^{^{3}}}[1-J_{0}(k_{t}\mathbf{r})]\alpha_{s}f(x,k_{t}^{2}),
\end{eqnarray}
where the integrated gluon distribution ($xg(x,\mu_{r}^{2})$) is
defined through the unintegrated gluon distribution
($f(x,k_{t}^{2})$) by
\begin{eqnarray}
xg(x,\mu_{r}^{2})&{\equiv}&\int^{\mu_{r}^{2}}\frac{dk_{t}^{2}}{k_{t}^{2}}f(x,k_{t}^{2}).
\end{eqnarray}
Indeed, the dipole cross section is directly connected via a
Fourier transform to the small-x UGD, whose evolution in $x$ is
regulated by the Balitsky- Fadin-Kuraev-Lipatov (BFKL) equation
[13]. Single and double BFKL pomeron exchanges have been
calculated by illustrating a dipole picture of high energy hard
scattering in the large $N_{c}$ limit in the leading logarithmic
approximation in Ref.[14].\\
The BFKL equation governs the evolution of the UGD,
 where the $k_{T}$-factorization is used in the high energy limit in which the QCD
interaction is described in terms of the quantity which depends on
the transverse momentum of the gluon. The gluon density in
inclusive and exclusive processes in a wide $Q^{2}$ region at low
$x$ is desirable, in the dominant double logarithmic (DLA)
contribution by the following form [15]
\begin{eqnarray}
xg(x,\mu_{r}^2){\propto}\exp{\bigg{[}
\frac{16N_{c}}{\beta_{0}}\ln\frac{x_{0}}{x}\ln\frac{t}{t_{0}}\bigg{]}},
\end{eqnarray}
where
$\frac{t}{t_{0}}{\equiv}\ln(\frac{\mu_{r}^2}{\Lambda^2_{QCD}})/\ln(\frac{Q_{0}^2}{\Lambda^2_{QCD}})$
and $\beta_{0}=11-\frac{2}{3}n_{f}$. Here $n_{f}$ is the number of
active flavours. The hard scale $\mu_{r}$ is assumed to have the
form $\mu_{r}^2=C/r^2+\mu_{0}^{2}$ where the parameters $C$ and
$\mu_{0}$ are obtained from the fit to the DIS data. A matching
between the dipole model gluon distribution and the collinear
approach, in the improved saturation model, is obtained [15,16] by
using a leading order gluon anomalous dimension $\gamma_{gg}$ as
\begin{eqnarray}
xg(x,\mu_{r}^2){\propto}I_{0}\bigg{(}2\sqrt{\frac{12}{\beta_{0}}\ln\frac{x_{0}}{x}\ln\frac{t}{t_{0}}}\bigg{)}
\exp \bigg{[} -\delta \ln\frac{t}{t_{0}}\bigg{]},
\end{eqnarray}
where $\delta=(11+\frac{2n_{f}}{27})/\beta_{0}$.\\
A novel formulation of the UGD for DIS in a way that accounts for
the leading powers in both the Regge and Bjorken limits is
presented in Ref.[17]. In this way, the UGD is defined by an
explicit dependence on the longitudinal momentum fraction $x$
which entirely spans both the dipole operator and the gluonic
Parton Distribution Function. The object of the BFKL evolution
equation at very small $x$ is the differential gluon structure
function of proton
\begin{eqnarray}
f(x,k_{t}^{2})&=&\frac{\partial{[xg(x,\mu_{r}^{2})]}}{\partial{\ln}\mu_{r}^{2}}|_{\mu_{r}^{2}=k^{2}_{t}}
\end{eqnarray}
which emerges in the color dipole picture (CDP) of inclusive deep
inelastic scattering (DIS)  and diffractive DIS into dijets [18].
Here $x$ and $k^{2}_{t}$ being the fractional momentum of proton
carried by gluon and the transverse momentum of gluon
respectively. Unintegrated distributions are required to describe
measurements where transverse momenta are exposed explicitly.
Eq.(6) cannot remain true as $x$ increases or decreases [19],
therefore modify Eq.(6), with the Sudakov form factor, to the form
\footnote{The Sudakov form factor can be defined into the dipole
models with the help of the following formula [20] :
$$
\sigma_{\mathrm{dip}}(x,r,Q^2)=\int^{r}_{0}dr'
r'\log(\frac{r}{r'})e^{-S(r',Q^2)}\nabla^{2}_{r'}\sigma_{\mathrm{dip}}(x,r')
$$}
[21]
\begin{eqnarray}
f(x,k_{t}^{2})&=&\frac{\partial{[xg(x,\mu_{r}^{2})}T(r,\mu_{r}^2)]}{\partial{\ln}\mu_{r}^{2}}|_{\mu_{r}^{2}=k^{2}_{t}},
\end{eqnarray}
with $T(r,\mu_{r}^2)=\exp(-S(r,\mu_{r}^2)))$ where the
perturbative Sudakov factor in the leading-order [22], for the
case of running coupling
$\alpha_{s}(\mu_{r}^2)=1/(b_{0}\ln\frac{\mu_{r}^2}{\Lambda_{\mathrm{QCD}}^2})$,
reads
\begin{eqnarray}
S^{(1)}_{\mathrm{pert}}(r,Q^2)=\frac{C_{A}}{2{\pi}b_{0}}\bigg{[}-\ln{\bigg{(}}\frac{Q^2}{\mu_{b}^2}{\bigg{)}}
+{\bigg{(}}
\frac{1+\alpha_{s}(\mu^2_{b})b_{0}\ln{\bigg{(}}\frac{Q^2}{\mu_{b}^2}{\bigg{)}}}{\alpha_{s}(\mu^2_{b})b_{0}}
{\bigg{)}}\ln{\bigg{(}}1+\alpha_{s}(\mu^2_{b})b_{0}\ln{\bigg{(}}\frac{Q^2}{\mu_{b}^2}{\bigg{)}}{\bigg{)}}\bigg{]},
\end{eqnarray}
where $b_{0}=\frac{11C_{A}-2n_{f}}{12}$ and
$\mu_{b}=2e^{-\gamma_{E}}/r$ where $\gamma_{E}{\approx}0.577$ is
the Euler-Mascheroni constant.\\
The proton structure function $F_{2}$ corresponds to the dipole
picture of DIS at small $x$, by Eq.(1), as
\begin{eqnarray}
F_{2}=F_{T}+F_{L}=\frac{Q^2}{4\pi^2\alpha_{em}}\bigg{(}\sigma_{T}^{\gamma^{*}p}+
\sigma_{L}^{\gamma^{*}p}\bigg{)}.
\end{eqnarray}
Since the photon wave function depends on mass of the quarks in
the color dipole model [23], then the light and heavy structure
functions are defined by the following form
\begin{eqnarray}
F_{T,L}=F^{l}_{T,L}+F^{h}_{T,L},
\end{eqnarray}
where $F^{l}_{T,L}$ is the sum of the contributions from the light
quark pairs, while $F^{h}_{T,L}$ is the contribution from the
heavy quarks ($c\overline{c}$, $b\overline{b}$ and may be
$t\overline{t}$~\footnote{The high $ep$ cms energy at the LHeC
will lead to the copious production of single top-quarks, about
$2{\times}10^{6}$ single top and $5{\times}10^{4}$ $t\overline{t}$
events [24]. }). So the Bjorken variable $x$ can be modified in
the gluon distribution and dipole cross section by the following
form
\begin{eqnarray}
x{\rightarrow}\widetilde{x}_{f}=\frac{Q^2+4m^2_{f}}{Q^2+W^2},
\end{eqnarray}
where $W^2$ is an invariant energy squared of the $\gamma^*p$
system and $m_{f}$ is the mass of the quark of flavour $f$.\\
Heavy-quarks production, in neutral current (NC) deep inelastic
electron-proton scattering (DIS) at HERA,  is the most important
quantum chromodynamics (QCD) tests. The production of heavy quarks
at HERA depends on the mass of these quarks and thus the
calculations of cross sections depend on a wide range of
perturbative scales $\mu^{2}$. The massive fixed-flavour-number
scheme (FFNS) [25] and the variable-flavour-number scheme (VFNS)
[26] are different approaches for  considering heavy quarks. FFNS
can be used on the threshold of $\mu^{2}{\approx}m_{f}^{2}$ and
for $\mu^{2}{\gg}m_{f}^{2}$ VFNS is used where the treatment of
resummation of collinear logarithms $\ln(\mu^{2}/m_{f}^{2})$ is
achieved. A general-mass variable-flavour-number scheme (GM-VFNS)
for calculation of the contributions of heavy quarks introduced in
Ref.[27]. For realistic kinematics it has to be extended to the
case of a GM-VFNS which is defined similarly to the zero-mass VFNS
(ZM-VFNS) in the $Q^{2}/m_{f}^{2}{\rightarrow}\infty$ limit [28].
In GM-VFNS the transition, from $n_{f}$ active flavors to
$n_{f}+1$, is considered in the construction of the charm-quark
parton distribution function. At some rather large scales (i.e.,
$Q^{2}>m_{f}^{2}$) the transition to two massive quarks (i.e.,
$n_{f}{\rightarrow}n_{f}+2$) has been discussed in Refs.[29,30].
In the GM-VFNS at high $Q^{2}$, the heavy-flavor structure
functions depend on the active flavor number since here $n_{f}=4$
for $m_{c}^{2}<\mu^{2}<m_{b}^{2}$, $n_{f}=5$ for
$m_{b}^{2}<\mu^{2}<m_{t}^{2}$ and $n_{f}=6$ for
$\mu^{2}{\geq}m_{t}^{2}$ is chosen.\\
The dynamics of flavor-singlet quark and gluon distribution
functions, $q^{s}$ and $g$, are defined by
\begin{eqnarray}
q^{s}(x,n_{f},\mu^{2})&=&\sum_{l=1}^{n_{f}}[f_{l}(x,n_{f},\mu^{2})+\overline{f}_{l}(x,n_{f},\mu^{2})],\nonumber\\
g(x,n_{f},\mu^{2})&=&f_{g}(x,n_{f},\mu^{2}).
\end{eqnarray}
The heavy-quark structure functions derived using the zero-mass
VFN scheme (ZMVFN) by the following form
\begin{eqnarray}
F^{ZMVFN}_{k}=\sum_{j=0}^{\infty}a_{s}^{j}(n_{f}+1)\sum_{i=q,g,h}
C_{k,i}^{(j)}(n_{f}+1){\otimes}f_{i}(n_{f}+1)
\end{eqnarray}
where $C^{,}s $ are the Wilson coefficients at the $j$-th order
and $k=2$ and $L$ and the $\otimes$ symbol denotes the convolution
integral which turns into a simple multiplication in Mellin
$N$-space. The notation is defined by $a(x)\otimes
b(x)=\int_{x}^{1}\frac{dz}{z}a(z)b(\frac{x}{z})$. Here
$a_{s}=\frac{\alpha_{s}}{4\pi}$ is the QCD running coupling.
Eq.(13), at asymptotically large momentum transfer
$Q^{2}{\gg}m_{f}^{2}$, is valid. For $Q^{2}{\simeq}m_{f}^{2}$ VFNS
is valid which it includes a combination of the ZMVFN with FFNS.
In this case the heavy- quark structure functions are
\begin{eqnarray}
F^{FFNS} _{k}=\sum_{j=0}^{\infty}a_{s}^{j}(n_{f})\sum_{i=q,g}
H_{k,i}^{(j)}(n_{f}){\otimes}f_{i}(n_{f}),
\end{eqnarray}
where $H^{,}s$ are the Wilson coefficients for the DIS heavy-quark
production [30].\\
In this paper we present the heavy quark structure functions due
to the dipole models in the collinear approach. These calculations
are based on the generalized double asymptotic scaling (DAS)
approach [31-34]. We continue our investigations and analyze the
heavy quark structure functions and those ratios in a wide range
of $r$ in section II. In this section, the heavy quark structure
functions  can be combined with the Sudakov form factor. Sections
III and IV
contains our results and conclusions respectively.\\

\subsection{II. Method}

$\bullet~ \mathrm{\mathbf{Structure~ Functions:}}$\\

The heavy quark structure functions in DIS in ep colliders  are
obtained from the measurements of the inclusive heavy quark cross
sections, which will be an important test of the QCD in the LHeC
and FCC-he colliders [24]. The reduced cross section of the top
quark is defined in terms of the top structure functions by the
following form:
\begin{eqnarray}
\sigma^{{h}\overline{{h}}}_{\mathrm{red}}(x,Q^{2})=
\frac{xQ^{4}}{2\pi
\alpha_{EM}^{2}[(1+(1-y)^{2}]}\frac{d^{2}\sigma^{{h}\overline{{h}}}}{dxdQ^{2}}=F_{2}^{h}(x,Q^{2})-f(y)F_{L}^{h}(x,Q^{2}),
\end{eqnarray}
where $f(y)=\frac{y^{2}}{1+(1-y)^{2}}$. In HERA kinematic range
the contribution $F_{L}^{h\overline{h}}$ is small. Therefore the
heavy-quark structure function $F_{2}^{h}$ is obtained from the
measured heavy-quark cross sections. The ratio
$R^{h}(x,Q^{2})=F_{L}^{h}(x,Q^{2})/F_{2}^{h}(x,Q^{2})$ will extend
in future circular colliders (i.e., LHeC and FCC-he). Indeed,
these new colliders are the ideal place to resolve this ratio.\\
In the small $x$ range, where the gluon contribution is dominant,
the heavy quark structure functions in the collinear generalized
DAS approach are given by [34]
\begin{eqnarray}
F_{k}^{h}(x,\mu_{r}^{2}){\simeq}~e^{2}_{h}\sum_{n=0}(\frac{\alpha_{s}}{4\pi})^{n+1}B^{(n)}_{k,g}(x,\xi_{r})
{\otimes} xg(x,\mu_{r}^{2}),
\end{eqnarray}
where $B_{k,g}$ are the collinear Wilson coefficient functions in
the high energy regime [34] and $e^{2}_{h}$ is the squared charge
of the heavy flavor. Here, $n$ denotes the order in running
coupling $\alpha_{s}$ and $\xi_{r}=\frac{m_{f}^{2}}{\mu_{r}^{2}}$.
The explicit expressions for the coefficient functions at the
leading order (LO) up to next-to-next-to-leading order (NNLO)
approximations  are relegated in Appendix. The default
renormalisation and factorization scales are set to be equal
$\mu_{R}^{2}=\mu_{r}^{2}+4m_{f}^2$ and
$\mu_{F}^{2}=\mu_{r}^{2}$.\\
The integrated and unintegrated gluon distributions from the GBW
and BGK models are obtained in Ref.[35], which were formulated on
the position-space version of the $k_{t}$-factorization formula.
The gluon density is parametrized at the scale $\mu_{r}^2$ using
the running coupling $\alpha_{s}$ by the following form
\begin{eqnarray}
xg(x,\mu_{r}^2)=\frac{\sigma_{0}}{16\pi^3}Q_{0}^2(\frac{x_{0}}{x})^\lambda(11C_{A}-2n_{f})\ln\bigg{(}\frac{\mu_{r}^2}{\Lambda_{QCD}^{2}}
\bigg{)}.
\end{eqnarray}
where $C_{A}=N_{c}=3$ is the Casimir operator in the fundamental
and adjoint representation of the $\mathrm{SU(N_{c})}$ color group
and the QCD parameter $\Lambda$ is extracted by
$\alpha_{s}(M_{Z}^{2})$ using the c and b-quark
threshold\footnote{ In Refs.[36] and [37], the massive quarks in
NLO dipole factorization for DIS are considered. The NLO
corrections for the dipole factorization of DIS structure
functions at low $x$ is considered using light front perturbative
theory as
$$
|\gamma^{*}_{T,L}>_{NLO}=|\gamma^{*}_{T,L}>_{q\overline{q}}+|\gamma^{*}_{T,L}>_{q\overline{q}g}.
$$}.\\
The parameters of the model (i.e.,$\sigma_{0}$, $x_{0}$ and
$\lambda$ ) depend on the active flavor number are found from a
fit to small-$x$ data in Table I.
 \begin{table}
\centering \caption{The fixed parameters according to Ref.[23]
from the fit results to the HERA data using the dipole cross
section.
  }\label{table:table1}
\begin{minipage}{\linewidth}
\renewcommand{\thefootnote}{\thempfootnote}
\centering
\begin{tabular}{|l|c|c|c|c|c|c|c||} \hline\noalign{\smallskip} Ref. & $m_{l}$[GeV]  &
$m_{c}$[GeV] & $m_{b}$[GeV] & $\sigma_{0}[mb]$ & $\lambda$  & $x_{0}/10^{-4}$ & $\chi^{2}/N_{\mathrm{dof}}$\\
\hline\noalign{\smallskip}
[23] & 0.14 & 1.4 & - & 27.32$\pm$0.35 & 0.248$\pm$0.002 & 0.42$\pm$0.04 & 1.60  \\
\hline\noalign{\smallskip}
[23] & 0.14 & 1.4 & 4.6 & 27.43$\pm$0.35 & 0.248$\pm$0.002 & 0.40$\pm$0.04 & 1.61  \\
\hline\noalign{\smallskip}
\end{tabular}
\end{minipage}
\end{table}
After exploiting the low $x$ behavior of the gluon density (i.e.,
Eq.(17)), Eq.(16) can be rewritten as
\begin{eqnarray}
F_{k}^{h}(x,\mu_{r}^{2})&{\simeq}&M^{h}_{k,g}(x,\mu_{r}^{2},\lambda)xg(x,\mu_{r}^{2})
\end{eqnarray}
where
\begin{eqnarray}
M^{h}_{k,g}(x,\mu_{r}^{2},\lambda)=e^{2}_{h}\sum_{n=0}(\frac{\alpha_{s}}{4\pi})^{n+1}\int_{x}^{x_{2}}
B^{(n)}_{k,g}(y,\xi_{r})y^{\lambda-1}dy.
\end{eqnarray}
Therefore, the explicit form of the heavy structure functions at
the LO approximation, in the particular case of off-shell initial
gluons (when $k^2=0$) is
\begin{eqnarray}
F_{2}^{h}(x,\mu_{r}^{2})&=&e^{2}_{h}\frac{3\sigma_{0}}{16\pi^3}Q_{0}^2(\frac{x_{0}}{x})^\lambda\int_{x}^{x_{2}}
\bigg{\{} -2y\beta\bigg{[}1-4y(2-\xi_{r})(1-y)-\bigg{(}
1-2y(1-2\xi_{r})\nonumber\\
&&+2y^2(1-6\xi_{r}-4\xi^2_{r}) \bigg{)}L(\beta) \bigg{]} \bigg{\}}
y^{\lambda-1}dy,\nonumber\\
F_{L}^{h}(x,\mu_{r}^{2})&=&e^{2}_{h}\frac{3\sigma_{0}}{16\pi^3}Q_{0}^2(\frac{x_{0}}{x})^\lambda\int_{x}^{x_{2}}
8y^2\beta\bigg{[}(1-y)-2y\xi_{r}L(\beta)\bigg{]}y^{\lambda-1}dy,
\end{eqnarray}
where $\beta=\sqrt{1-\frac{4x\xi_{r}}{1-x}}$ and
$L(\beta)=\frac{1}{\beta}\ln\frac{1+\beta}{1-\beta}$. The ratio
$R^{h}(x,\mu_{r}^{2})=\frac{F_{L}^{h}(x,\mu_{r}^{2})}{F_{2}^{h}(x,\mu_{r}^{2})}$
can be presented as
\begin{eqnarray}
R^{h}(x,\mu_{r}^{2})&=&\frac{M_{L,g}(x,\mu_{r}^{2},\lambda)}{M_{2,g}(x,\mu_{r}^{2},\lambda))}\nonumber\\
&&=\frac{\int_{x}^{x_{2}}
8y^2\beta{[}(1-y)-2y\xi_{r}L(\beta){]}y^{\lambda-1}dy}{\int_{x}^{x_{2}}
{\{} -2y\beta{[}1-4y(2-\xi_{r})(1-y)-{(}
1-2y(1-2\xi_{r})+2y^2(1-6\xi_{r}-4\xi^2_{r}){)}L(\beta){]}{\}}y^{\lambda-1}dy}
\end{eqnarray}
where the ratio is independent of the gluon density, the Sudakov
form factor and the running coupling at the LO approximation. The
Sudakov form factor can be included by using Eq.(8) and
generalizing it to the heavy quark structure functions (i.e.,
Eq.(18)) by the following form
\begin{eqnarray}
F_{k}^{h}(x,\mu_{r}^{2})=e^{-S(r,\mu_{r}^2)}M^{h}_{k,g}(x,\mu_{r}^{2},\lambda)xg(x,\mu_{r}^{2})
\end{eqnarray}
The heavy quark structure functions now depend on the non-linear
gluon evolution at small $x$ due to the Sudakov effects, which
become relevant for processes with two distinct scales.\\

$\bullet~ \mathrm{\mathbf{Bounds:}}$\\

In the following, we discuss further bounds [7, 38-41] for
$F_{2}^{c}/F_{2}$ and $F_{2}^{b}/F_{2}$ which follow from the
standard dipole picture. Indeed, we give correlated bounds for
$F_{2}^{c,b}/F_{2}$ versus $F_{L}/F_{2}$ where the higher Fock
components of the photon wave function affect these bounds. In
Refs.[38,39], the authors have shown that the upper bound is
independent of $Q^2$ and numerically leads to $
\frac{F_{L}(W,Q^2)}{F_{2}(W,Q^2)}{\leq}g_{\mathrm{max}}=0.27139 $
for the case of massless quarks. A stronger bound can obtained by
considering the effect of the charm and bottom quarks on the ratio
\begin{eqnarray}
\frac{F_{L}^{\mathrm{light+c+b}}}{F_{2}^{\mathrm{light+c+b}}}
=\frac{F_{L}+F_{L}^{\mathrm{c}}+F_{L}^{\mathrm{b}}}{F_{2}+F_{2}^{\mathrm{c}}+F_{2}^{\mathrm{b}}}
=\frac{F_{L}/F_{2}+F_{L}^{\mathrm{c}}/F_{2}^{\mathrm{c}}F_{2}^{\mathrm{c}}/F_{2}+F_{L}^{\mathrm{b}}/F_{2}^{\mathrm{b}}F_{2}^{\mathrm{b}}/F_{2}}
{1+F_{2}^{\mathrm{c}}/F_{2}+F_{2}^{\mathrm{b}}/F_{2}}\nonumber\\
{\leq}g_{\mathrm{max}}\frac{1+g^{c}_{\mathrm{max}}F_{2}^{\mathrm{c}}/F_{2}+g^{b}_{\mathrm{max}}F_{2}^{\mathrm{b}}/F_{2}}
{1+F_{2}^{\mathrm{c}}/F_{2}+F_{2}^{\mathrm{b}}/F_{2}}{\leq}g_{\mathrm{max}}
\end{eqnarray}
In this case the bound on the ratio $F_{L}/F_{2}$ will depend on
the values of $F^{c}_{2}/F_{2}$ and  $F^{b}_{2}/F_{2}$, where
these bounds (i.e., $F^{h}_{2}/F_{2}$ ) can further restrict the
kinematical range of applicability of the dipole picture in future
colliders [24,42]. In the CDP, the gluon distribution has been
recently determined in Ref.[43] at low $x$ by the following form
\begin{eqnarray}
\alpha_{s}(\mu_{r}^2)xg(x,\mu_{r}^2)=\frac{9\pi}{R_{e^{+}e^{-}}}\frac{1}{2\rho+1}F_{2}(\eta_{L}x,\mu_{r}^2),
\end{eqnarray}
where $R_{e^{+}e^{-}}=N_{c}\sum_{f}e_{f}^{2}$ and
$\eta_{L}{\simeq}0.40$ is the rescaling factor. The $\rho$
parameter describes the ratio of the average transverse momenta
$\rho=\frac{<\overrightarrow{k}^{2}_{\bot}>_{L}}{<\overrightarrow{k}^{2}_{\bot}>_{T}}$,
which the transverse momentum $\overrightarrow{k}^{2}_{\bot}$ is
introduced into four momenta of the quark and antiquark. The
quantity of $\rho$, for $Q^2{\gg}\Lambda^{2}_{\mathrm{sat}}$, was
used to be $\rho=4/3$ [44]. The ratio of the longitudinal to the
transversal photoabsorption cross sections is given by
\begin{eqnarray}
R=\frac{\sigma_{L}^{\gamma^{*}p}}{\sigma_{T}^{\gamma^{*}p}}=\frac{1}{2\rho},
\end{eqnarray}
where factor 2 originates from the difference in the photon wave
functions. In terms of the proton structure functions, $F_{2}$ and
$F_{L}$, the ratio becomes\footnote{The colored sector of the
virtual photon wave-functions contains both $q\overline{q}$ and
$q\overline{q}g$ components at the NLO approximation. Expansion of
the structure functions, $F_{2}$ and $F_{L}$, in Fock state in the
CDM are given by
$$
F_{2,L}=F^{q\overline{q}}_{2,L}+F^{q\overline{q}g}_{2,L}+...
$$
where at higher Fock states one can be derived [37] the modified
CDM bound for the ratio $\frac{F_{L}}{F_{2}}$ as
$$
\bigg{(}\frac{F_{L}}{F_{2}}\bigg{)}_{\mathrm{NLO}}=\bigg{(}\frac{F_{L}}{F_{2}}\bigg{)}_{\mathrm{LO}}
\frac{1+\delta\epsilon}{1+\epsilon}
$$
where
$\epsilon=\frac{F_{2}^{q\overline{q}g}}{F_{2}^{q\overline{q}}}$
and $0{\leq}\delta{\leq}3.7$.}
\begin{eqnarray}
\frac{F_{L}}{F_{2}}=\frac{1}{1+2\rho}.
\end{eqnarray}
With imposing consistency between the CDP and the pQCD, the gluon
distribution function is obtained by expressing the proton
structure function in terms of $F_{L}$ as
\begin{eqnarray}
\alpha_{s}(\mu_{r}^2)xg(x,\mu_{r}^2)=\frac{3\pi}{\sum_{f}e_{f}^{2}}F_{L}(\eta_{L}x,\mu_{r}^2).
\end{eqnarray}
Therefore the ratio ${F_{2}^{h}}/{F_{2}}$ and
${F_{L}^{h}}/{F_{L}}$ are defined by
\begin{eqnarray}
\frac{F_{2}^{h}}{F_{2}}=\frac{3}{4}\frac{e_{h}^{2}}{\eta_{L}^{\lambda}\sum_{f}e_{f}^{2}}\frac{\int_{x}^{x_{2}}
B^{(0)}_{2,g}(y,\xi_{r})y^{\lambda-1}dy}{2\rho+1}
\end{eqnarray}
and
\begin{eqnarray}
\frac{F_{L}^{h}}{F_{L}}=\frac{3}{4}\frac{e_{h}^{2}}{\eta_{L}^{\lambda}\sum_{f}e_{f}^{2}}{\int_{x}^{x_{2}}
B^{(0)}_{L,g}(y,\xi_{r})y^{\lambda-1}dy}.
\end{eqnarray}
These ratios are very interesting in the range of available HERA
energy and its extension to future energies in LHeC and FCC-he
colliders.\\


\subsection{III. Numerical Results}

In the present paper we consider the heavy quark structure
functions to the deep inelastic proton structure function, which
are directly related with the gluon distribution of the proton in
the CDP approach at low $x$. Everywhere below, we set the charm
and bottom masses to be equal to $m_{c}=1.4~\mathrm{GeV}$ and
$m_{b}=4.6~\mathrm{GeV}$ according to Ref.[23]. In accordance with
the values recommended by the Higgs Cross Section Working Group
[45], the top-quark pole mass is set as in the NNPDF default
analysis to $m_{t}=172.5~\mathrm{GeV}$ [46]. To estimate the scale
uncertainties of our calculations, the standard variations in
default renormalization and factorization scales, which were set
to be equal to $\mu_{R}^{2}=\mu_{r}^{2}+4m^{2}$ and
$\mu_{F}^{2}=\mu_{r}^{2}$, respectively, were introduced.
 \begin{table}[h]
\centering \caption{The transverse separation range of $r$ in the
HERA and future colliders (i.e., LHeC and FCC-he) with the
inelasticity $y{\leq}1$ for $x=0.0013$ and 0.0050.
  }\label{table:table1}
\begin{minipage}{\linewidth}
\renewcommand{\thefootnote}{\thempfootnote}
\centering
\begin{tabular}{|l|c||c|c|c||} \hline\noalign{\smallskip} Collider &
$\sqrt{s}$[GeV] & x=0.0013 & x=0.0050 \\
\hline\noalign{\smallskip}
FCC-he & 3500 & r$>$0.005 & r$>$0.002  \\
\hline\noalign{\smallskip}
LHeC & 1300 & r$>$0.01 & r$>$0.006   \\
\hline\noalign{\smallskip}
HERA & 320 & r$>$0.05 & r$>$0.02  \\
\hline\noalign{\smallskip}
\end{tabular}
\end{minipage}
\end{table}
In recent years [47,48], the phenomenological various successful
methods have examined charm and bottom structure functions. This
importance, along with the t-quark density [49,50], can be
explored at future circular collider energies.\\
\begin{figure}[h]
\includegraphics[width=0.6\textwidth]{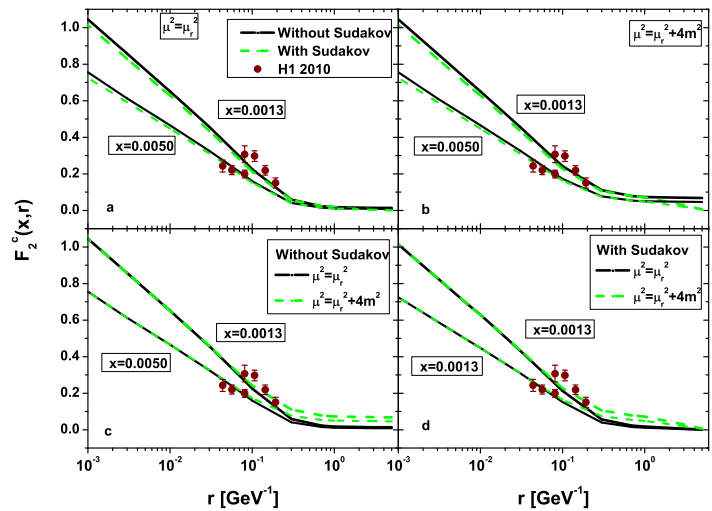}
\caption{Comparison of the H1 data from [51] for the charm
$F_{2}^{c}$ structure function with the results from unifying the
color dipole picture and double asymptotic scaling approaches with
the parameters in Table I in a wide range of the transverse
separation $\mathrm{r}~[\mathrm{GeV}^{-1}]$. The uncertainties are
due to $\mu^{2}=\mu_{r}^{2}+4m^{2}$, $\mu^{2}=\mu_{r}^{2}$ and the
Sudakov form factor with $x=0.0013$ and $0.0050$. }\label{Fig1}
\end{figure}
\begin{figure}[h]
\includegraphics[width=0.6\textwidth]{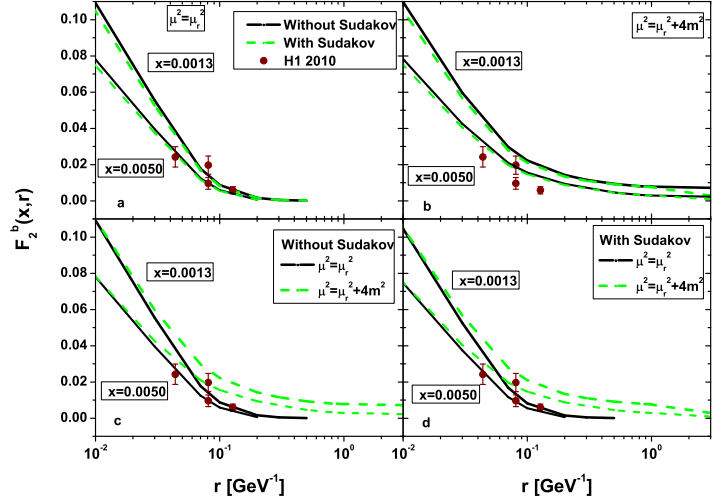}
\caption{The same as Fig.1 for the bottom structure
function.}\label{Fig2}
\end{figure}
Our numerical results for charm and bottom structure functions,
$F_{2}^{c}$ and $F_{2}^{b}$, are shown in Figs.1 and 2,
respectively, in comparison with the H1 data [51]. To estimate the
uncertainties of our calculations, the standard variations in
default scales (i.e., renormalization and factorization) and the
behavior of the Sudakov form factor are introduced. We observe
that the predictions obtained using  unifying the color dipole
picture and double asymptotic scaling approaches in a proton are
in perfect agreement with the H1 data in a wide range of $r$ for
$x=0.0013$ and $0.0050$ within the total experimental
uncertainties. These results for $F_{2}^{c}$ and $F_{2}^{b}$, in
Figs.1 and 2, increase as $r$ decreases. As a result, we predict
that at very low $r$, the charm and bottom structure functions
will increase at the FCC-he than the LHeC and HERA at high
inelasticity according to Table II. The uncertainties (without and
with Sudakov effects) increase as $r$ increase. In Fig.1-c,d, we
observe that the uncertainties for the charm structure functions
increase for $r{\gtrsim}2{\times}10^{-1}~\mathrm{GeV}^{-1}$ and
for the bottom structure functions increase for
$r{\gtrsim}3{\times}10^{-2}~\mathrm{GeV}^{-1}$ in Fig.2-c,d. The
effect of the Sudakov form factors in the charm and bottom
structure functions are shown in Figs.1-a,b and 2-a,b for the
renormalization and factorization scales, respectively.
Differences between the results (with and without Sudakov form
factor) for the charm and bottom structure functions are very
small and visible at small $r$. One can see that the Sudakov
factor mostly disappears in the large-$r$ region for
$\mu^{2}=\mu_{r}^{2}$ and survives for
$\mu^{2}=\mu_{r}^{2}+4m^{2}$ at $r>2~\mathrm{GeV}^{-1}$. The
changes are less obvious compared to the charm and bottom
structure functions without Sudakov form factor, which is because
the large-$r$ region, where the dipole cross section was affected
the most, is largely suppressed by the photon wave function. In
conclusion, the structure functions with Sudakov form factor seem
to show slightly more change in a wide range of $r$. We can add
these results as associated with the LHeC simulated uncertainties
[24]. These simulated uncertainties for $F_{2}^{c}$ and
$F_{2}^{b}$ measurements were recently published by the LHeC study
group
and reported by Ref.[24]\footnote{For further discussion, such predictions can be found in Figs. 3.4, 3.6 and 3.7 of Ref.[24].}.\\
In Figs.3 and 4, the importance of the longitudinal structure
function for  charm and bottom pair production, $F_{L}^{c}$ and
$F_{L}^{b}$, are examined according to Table II for colliders
(HERA, LHeC and FCC-he) in a wide range of $r$. The behavior of
these structure functions considers with and without Sudakov form
factor in a wide range of $r$ for $x=0.0013$ and $0.0050$ in
Figs.3 and 4.
\begin{figure}[h]
\includegraphics[width=0.6\textwidth]{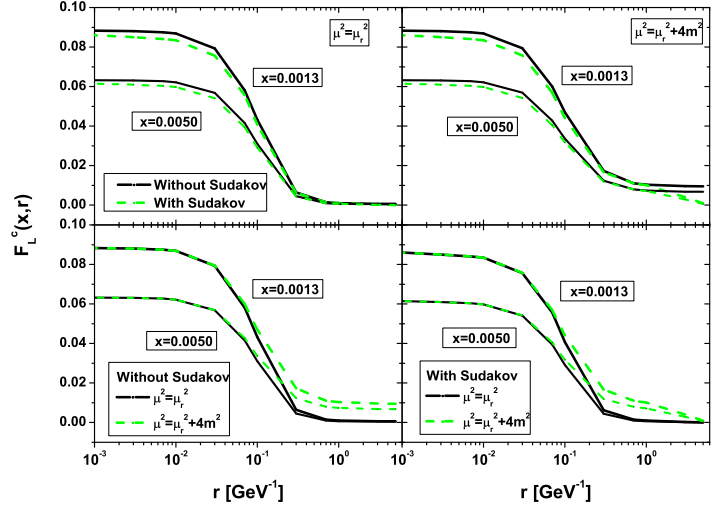}
\caption{The behavior of the charm $F_{L}^{c}$ structure function
due to unifying the color dipole picture and double asymptotic
scaling approaches with the parameters in Table I in a wide range
of the transverse separation $\mathrm{r}~[\mathrm{GeV}^{-1}]$. The
uncertainties are due to $\mu^{2}=\mu_{r}^{2}+4m^{2}$,
$\mu^{2}=\mu_{r}^{2}$ and the Sudakov form factor with $x=0.0013$
and $0.0050$.}\label{Fig3}
\end{figure}
\begin{figure}[h]
\includegraphics[width=0.6\textwidth]{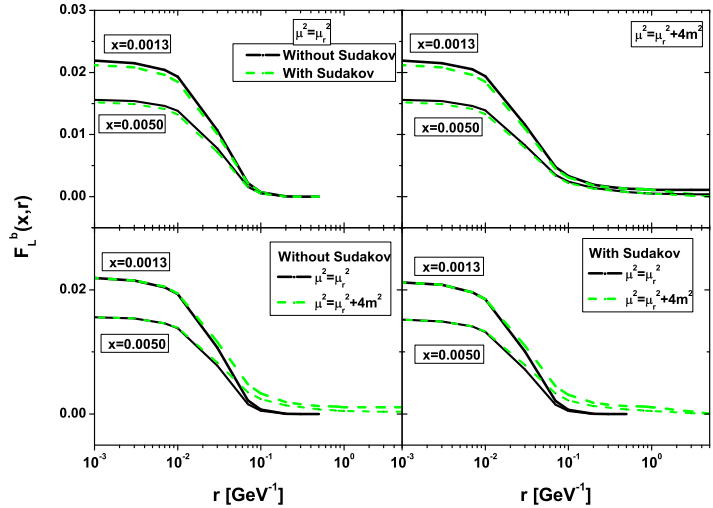}
\caption{The same as Fig.3 for the bottom structure
function.}\label{Fig4}
\end{figure}
In Fig.3-c,d, we observe that the uncertainties for the
$F_{L}^{c}$ increase for $r{\gtrsim}10^{-1}~\mathrm{GeV}^{-1}$ and
for the $F_{L}^{b}$ increase for
$r{\gtrsim}5{\times}10^{-2}~\mathrm{GeV}^{-1}$ in Fig.4-c,d. The
effect of the Sudakov form factors in the $F_{L}^{c}$ and
$F_{L}^{b}$ are shown in Figs.13-a,b and 4-a,b for the
renormalization and factorization scales, respectively.
Differences between the results (with and without Sudakov form
factor)  are very small and visible at small $r$. One can see that
the Sudakov factor mostly disappears in the large-$r$ region for
$\mu^{2}=\mu_{r}^{2}$ and survives for
$\mu^{2}=\mu_{r}^{2}+4m^{2}$ at $r>2~\mathrm{GeV}^{-1}$. According
to Figs.3 and 4, we observe that the longitudinal structure
function for charm and bottom increase as $r$ decreases due to the
HERA range of $r$ (see Table II). We observe that the $F_{L}^{c}$
and $F_{L}^{b}$ in the LHeC and FCC-he range energy makes the
transition from the large $r$ to the low $r$ forms. The
longitudinal structure functions for charm
($r{\gtrsim}3{\times}10^{-1}~\mathrm{GeV}^{-1}$) and bottom
($r{\gtrsim}10^{-1}~\mathrm{GeV}^{-1}$) are slowly varying  and
reach to zero for large $r$. These values have a constant rate at
lower $r$. The ratio of the longitudinal structure functions,
$\frac{F_{L}^{c}}{F_{L}^{b}}$, is of the
$\mathcal{O}(\frac{m_b}{m_c})$ order for
$r<10^{-2}~\mathrm{GeV}^{-1}$. This ratio shows that the
importance of measuring the longitudinal structure function for
bottom quark is not less than charm quark in the process analysis
of new colliders.\\
\begin{figure}[h]
\includegraphics[width=0.55\textwidth]{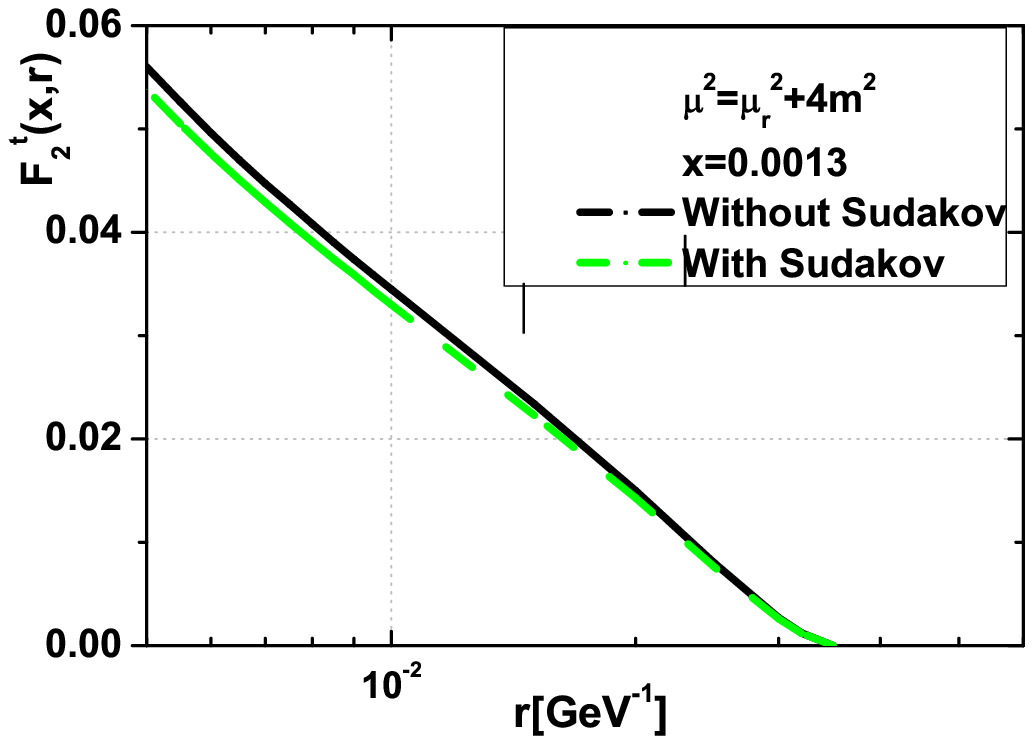}
\caption{Results of the top $F_{2}^{t}$ structure function with
the bottom parameters in Table I in a wide range of the transverse
separation $\mathrm{r}~[\mathrm{GeV}^{-1}]$ at
$\mu^{2}=\mu_{r}^{2}+4m^{2}$ with and without Sudakov form factor
for $x=0.0013$.}\label{Fig5}
\end{figure}
Considering the top structure function in unifying the color
dipole picture and double asymptotic scaling approaches is
interesting because the production of top quarks in
electron-proton collisions at LHeC and FCC-he can provide a
stringent test of new physics at ultra-high energy (UHE). In
Fig.5, the top structure function is predicted in a wide range of
the transverse separation range of $r$ with
$\mu^{2}=\mu_{r}^{2}+4m^{2}$ for $x=0.0013$. The Sudakov form
factor effect is compared with these results in Fig.5. The
difference between the results (with and without Sudakov form
factor) for the top structure function is very small and visible
at small $r$. One can see that the Sudakov factor mostly
disappears in the large-$r$ region. It is observed that in the
HERA energy range ($r{\gtrsim}0.05~\mathrm{GeV}^{-1}$), the top
structure function is zero. This probability increases as the
energy range increases to future colliders (especially FCC-he). It
is clear that the top structure function will increase at the
FCC-he than the LHeC at high inelasticity\footnote{Notice that the
large inelasticity is only for scattered electron energies much
smaller than the electron beam energy (i.e., $E'_{e}{\ll}E_{e}$
and $y=1-E'_{e}/E_{e}$). In this region where $E'_{e}$ is small,
the electromagnetic and hadronic backgrounds are important [24].},
according to Table II. It reaches $F_{2}^{t}\simeq~0.05$ at
$r{\simeq}0.005~\mathrm{GeV}^{-1}$ in the FCC-he energy range and
$F_{2}^{t}\simeq~0.03$ at $r{\simeq}0.01~\mathrm{GeV}^{-1}$ in the
LHeC energy range for $x=0.0013$ with $\mu^{2}=\mu_{r}^{2}+4m^{2}$.\\
\begin{figure}[h]
\includegraphics[width=0.55\textwidth]{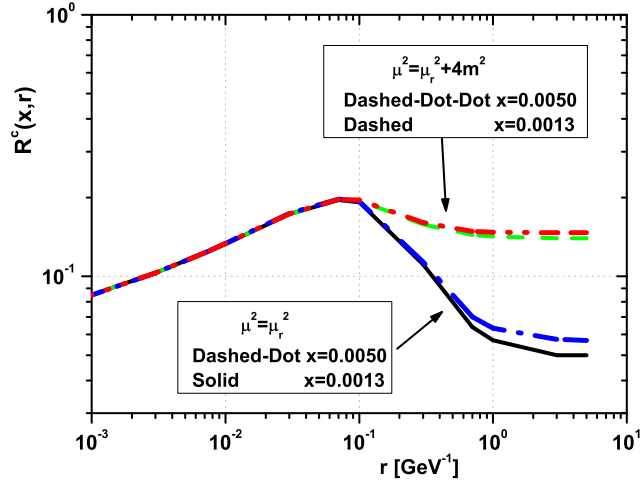}
\caption{$R^{c}$ evaluated as a function of $r$ with
$\mu^2=\mu_{r}^2$ and  $\mu^2=\mu_{r}^2+4m^{2}$ for $x=0.0013$ and
$0.0050$.}\label{Fig6}
\end{figure}
\begin{figure}[h]
\includegraphics[width=0.55\textwidth]{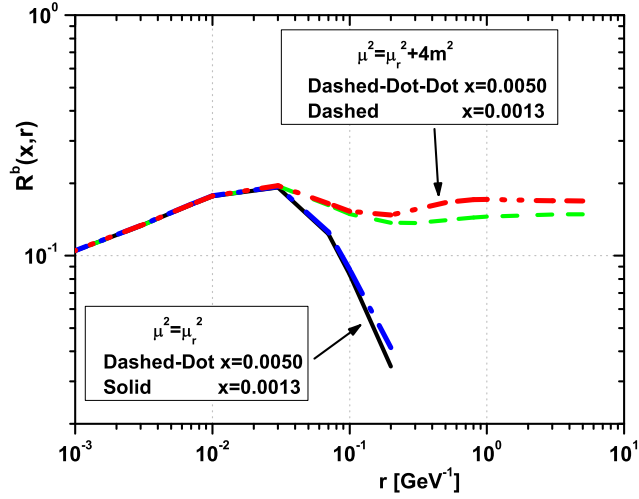}
\caption{The same as Fig.5 for the ratio of the bottom structure
functions $R^{b}$.}\label{Fig7}
\end{figure}
Results of our calculations for
$R^{c}=\frac{F_{L}^{c}}{F_{2}^{c}}$ and
$R^{b}=\frac{F_{L}^{b}}{F_{2}^{b}}$ are presented in Fig.6 and 7
respectively, where we plot these ratios as a function of $x$ in a
wide $r$ range with $\mu^2=\mu_{r}^2$ and
$\mu^2=\mu_{r}^2+4m^{2}$. We observe from Figs.6 and 7 that these
results for $r{\gtrsim}10^{-1}~\mathrm{GeV}^{-1}$ leads to a flat
behavior of $R^{c}$ and $R^{b}$ with $\mu^2=\mu_{r}^2+4m^2$ and
decrease sharply with $\mu^2=\mu_{r}^2$. The results obtained with
the renormalization and factorization scales for $R^c$ and $R^b$
are compatible at $r<10^{-1}~\mathrm{GeV}^{-1}$ and
$3{\times}10^{-2}~\mathrm{GeV}^{-1}$ respectively and have the
largest uncertainties at $r>10^{-1}~\mathrm{GeV}^{-1}$ and
$3{\times}10^{-2}~\mathrm{GeV}^{-1}$. Our calculations show an
$x$-independent behavior of $R^c$ and $R^b$ in a wide range of $r$
with the renormalization and factorization scales. For larger
values of $r$, some dependence on $x$ appears, especially in
$R^{c}$ with $\mu^2=\mu_{r}^2$ and in $R^{b}$ with
$\mu^2=\mu_{r}^2+4m^2$. The maximum value of $R(x,r)$ is equal to
$\simeq 0.2$ for charm and bottom ratios at $r\simeq
10^{-1}~\mathrm{GeV}^{-1}$ and
$3{\times}10^{-2}~\mathrm{GeV}^{-1}$ respectively. We observe that
the maximum value shifts to smaller values of $r$ for the bottom
quark than the charm. These results are comparable with others in literature [24, 31, 32, 34, 52].\\
\begin{figure}[h]
\includegraphics[width=0.55\textwidth]{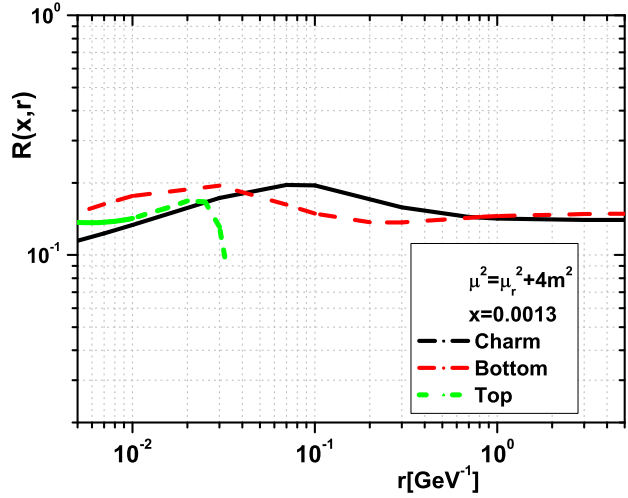}
\caption{$R^{c}, R^{b}$ and $R^{t}$ evaluated as a function of $r$
with $\mu^2=\mu_{r}^2+4m^{2}$ for $x=0.0013$.}\label{Fig8}
\end{figure}
In order to assess the significance of the ratio of structure
functions in a wide range of the collider energies (from HERA
until FCC-he), we show in Fig.8 the $r$ dependences of $R^{c}$,
$R^{b}$ and $R^{t}$ evaluated with $\mu^2=\mu_{r}^2+4m^{2}$ for
$x=0.0013$. We observe from Fig.8 that the charm and bottom
predictions have similar behaviors in a wide range of $r$ and
collider energies (according to Table II). The charm and bottom
ratios increase until $r\simeq 10^{-1}~\mathrm{GeV}^{-1}$ and
$3{\times}10^{-2}~\mathrm{GeV}^{-1}$ respectively, then decrease
and have a flat ($r$-independent) behavior for large values of $r$
($r{\geq}0.3~\mathrm{GeV}^{-1}$). In Fig.8, we observe the ratio
of the top structure functions according to the FCC-he
center-of-mass energy in the inelasticity range $0<y<1$ due to the
coefficients in Table I. It continues to rise with $r$, then fall
after reaching a maximum\footnote{For further discussion please
see Refs.[34,53]}. Such results seem to be extremely important for
future experiments, in particular, for experiments at the LHeC and
FCC-he.\\
\begin{figure}[h]
\includegraphics[width=0.55\textwidth]{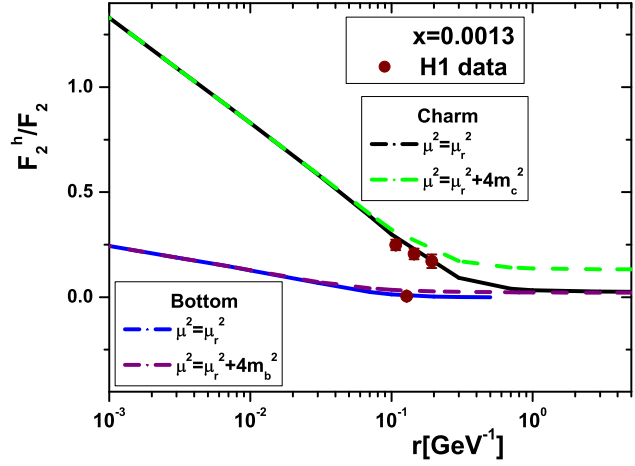}
\caption{Ratios $F_{2}^{c}/F_{2}$ and $F_{2}^{b}/F_{2}$ as
functions of $r$ with $\mu^2=\mu_{r}^2$ and $\mu^2=\mu_{r}^2+4m^2$
for x=0.0013. Experimental data are from the H1-Collaboration
[51,54].}\label{Fig9}
\end{figure}
\begin{table}[h]
\centering \caption{The ratios $F_{2}^{c}/F_{2}$ and
$F_{2}^{b}/F_{2}$ with the renormalization and factorization
scales are compared with the predictions of Ref.[7] from the BGK
and IP-sat models.  }\label{table:table1}
\begin{minipage}{\linewidth}
\renewcommand{\thefootnote}{\thempfootnote}
\centering
\begin{tabular}{|l|c||c|c||c|c||} \hline\noalign{\smallskip} x &
$Q^2[\mathrm{GeV}^2]$ & $\frac{F_{2}^{c}}{F2}|_{\mathbb{IP}-\mathrm{sat}..\mathrm{BGK}}$ & $\frac{F_{2}^{b}}{F2}|_{\mathbb{IP}-\mathrm{sat}..\mathrm{BGK}}$  &$\frac{F_{2}^{c}}{F2}|_{\mu_{r}^2..\mu_{r}^2+4m^2}$  & $\frac{F_{2}^{b}}{F2}|_{\mu_{r}^2..\mu_{r}^2+4m^2}$  \\
\hline\noalign{\smallskip}
$10^{-2}$ & 5 & 0.096-0.100 & 0.00042-0.00044 & 0.072-0.156 & $<$0.014  \\
          & 10 & 0.144-0.149 & 0.00165-0.00168 & 0.133-0.196 & 0.00008-0.02150  \\
          & 50 & 0.233-0.234 & 0.0115-0.0112 & 0.323-0.342 & 0.0151-0.0346  \\
\hline\noalign{\smallskip}
$10^{-4}$ & 5 & 0.150-0.154 & 0.0034-0.0033 & 0.080-0.165 & 0.0013-0.0274  \\
          & 10 & 0.197-0.200 & 0.0060-0.0057 & 0.139-0.202 & 0.0031-0.0286  \\
          & 50 & 0.280-0.280 & 0.0195-0.0186 & 0.328-0.346 & 0.0168-0.0367  \\
          \hline\noalign{\smallskip}
$10^{-6}$ & 5 & 0.184-0.194 & 0.0057-0.0053 & 0.080-0.165 & 0.0014-0.0275  \\
          & 10 & 0.230-0.238 & 0.0089-0.0086 & 0.139-0.202 & 0.0032-0.0287  \\
          & 50 & 0.305-0.308 & 0.0244-0.0235 & 0.328-0.346 & 0.0168-0.0367  \\
\hline\noalign{\smallskip}
\end{tabular}
\end{minipage}
\end{table}
In Fig.9, we plot ratios $F_{2}^{c}/F_{2}$ and $F_{2}^{b}/F_{2}$
as functions of $r$ with the renormalization and factorization
scales for x=0.0013. In this figure, the ratio of the structure
functions are compared with the H1 Collaboration data in Refs.[51]
and [54]. The error bars of the ratio ${F^{h}_{2}}/{F_{2}}$ are
determined by
$\Delta({F^{h}_{2}}/{F_{2}})={F^{h}_{2}}/{F_{2}}\sqrt{({{\Delta}F^{h}_{2}}/{F^{h}_{2}})^2+({{\Delta}F_{2}}/{F_{2}})^2}$,
where $\Delta{F^{h}_{2}}$ and $\Delta{F_{2}}$ are collected from
the H1 experimental data in Refs.[51] and [54] respectively. The
results obtained from the ratios are comparable to the H1 data
[51,54]. Realistic values of ${F^{c}_{2}}/{F_{2}}$ can only range
from zero to at most about 0.4 in the HERA energy range [38,
55]\footnote{The average value of the ratio ${F^{c}_{2}}/{F_{2}}$
is determined to be
$<{F^{c}_{2}}/{F_{2}}>=0.237{\pm}0.021^{+0.043}_{-0.039}$ in
Ref.[56].}. The results for the ratios ${F^{c}_{2}}/{F_{2}}$ and
${F^{b}_{2}}/{F_{2}}$, in Fig.9, are predicted at low values of
$r$ according to the LHeC and FCC-he energy range and will be able
to be considered in these collisions.  In particular, the
unphysical upper bound [38] ${F^{c}_{2}}/{F_{2}}=1$ will be
obtained at the low value of $r$ due to the FCC-he center-of-mass energy.\\
Recently, the structure functions $F_{2}, F_{L}$ and heavy quark
structure functions, $F^{c}_{2}, F^{b}_{2}$ from the models BGK
and IP-sat are predicted\footnote{Within the color dipole
approach, the impact parameter saturation model (IP-SAT) and the
BGK model include DGLAP evolution as the gluon density is
parametrized, in both models, at the initial scale $Q_{0}^{2}$
then scales $\mu^2$ by using the LO or NLO evolution equations.}
in the range $(x,Q^2):(10^{-6}-10^{-2},5.0-50~\mathrm{GeV}^2)$ in
Ref.[7]. We compared the ratios with the results of the BGK and
IP-sat models in Table III. One shows in this table that our
calculations are comparable with the predictions from the BGK and
IP-sat models. We can see that the predictions from the BGK and
IP-sat models lie between ($\mu_{r}^2{\lesssim} \mathrm{BGK},
\mathrm{IP-sat}{\lesssim}\mu_{r}^2+4m^2$) the bounds as the
maximum is of the order $(\mu_{r}^2)$-$(\mu_{r}^2+4m^2)$. The
differences between the results are due to the free-fit parameters
of the models in
Refs.[4] and [23].\\

\subsection{IV. Conclusions}

In this work we have computed the heavy quark structure functions
$F_{2,L}^{h} ,h=c,b,t$ within the $k_{t}$ factorization framework,
using unifying the color dipole picture and double asymptotic
scaling approaches for the integrated gluon density using the GBW
and BGK models at small Bjorken $x$ values. We have first
considered the structure functions $F_{2,L}^{h}$ in a wide range
of the transverse separation $r$ from the HERA to FCC-he
center-of-mass energy. Then we have obtained bounds on
$F_{L}^{h}/F_{2}^{h}$ as well as a correlated bound on the ratios
$F_{2}^{c}/F_{2}$ and $F_{2}^{b}/F_{2}$ as they are consistent
with the experimental data from HERA collider at moderate and
large $r$. It will be interesting to compare these bounds with
future results from measurements of these structure functions as
$r$ decreases.\\
We achieved a good agreement between the HERA experimental data
for the charm and bottom structure functions and our theoretical
predictions with the renormalization and factorization scales. We
demonstrated the importance of the contributions of $F_{L}^{c}$
and $F_{L}^b$ at small $r$ in further colliders. For the top quark
pair production, which will be one kind of important production
channel at LHeC and FCC-eh, the ratio of structure functions
(i.e., $R^{t}$) is determined and compared with the charm and
bottom ratios (i.e., $R^{c}$ and $R^{b}$) at small $r$ which are
dominated by the center-of-mass energies in new colliders at the
renormalization scale $\mu^2=\mu_{r}^2+4m^2$. To estimate the
uncertainties of our calculations, the standard variations in
default scales (i.e., renormalization and factorization) are
introduced. The uncertainty range of scales increases as $r$
increases.\\
Additionally, effects of the Sudakov form factor were investigated
for the heavy quark structure functions in a wide range of $r$.
The Sudakov form factor modifies the heavy quark structure
functions in the small region of $r$ owing to the saturation
effect. The effect is visible for a small value of $r$ and
disappears when $r$ increases. Moreover, we compared our
predictions of the ratio $F_{2}^{c}/F_{2}$ and $F_{2}^{b}/F_{2}$
with the BGK and IP-sat models at low values of $x$ and found all
good agreement with data sets  in the intervales of the
factorization and renormalization scales. We hope that this paper
at low $x$ and low $r$ will be useful in future phenomenological
studies of the heavy quark structure functions at future colliders
such as EIC, LHeC
and the FCC-he.\\

\subsection{ACKNOWLEDGMENTS}
The author is grateful to Razi University for the financial
 support of this project.\\


\subsection{APPENDIX }

In the high energy regime, defined by $x{\ll}1$, the coefficient
functions have the compact forms [34]
\begin{eqnarray}
B_{2,g}^{(0)}(1,\xi_{r})&=&
\frac{2}{3}[1+2(1-\xi_{r})J(\xi_{r})],\nonumber\\
B_{L,g}^{(0)}(1,\xi_{r})&=&\frac{4}{3}x_{2}\{1+6\xi_{r}-4\xi_{r}[1+3\xi_{r}]J(\xi_{r})\},\nonumber\\
B_{k,g}^{(1)}(x,\xi_{r})&=&\beta[R_{k,g}^{(1)}(1,\xi_{r})+4C_{A}^{2}B_{k,g}^{(0)}(1,\xi_{r})L_{\mu}],\nonumber\\
B_{k,g}^{(2)}(x,\xi_{r})&=&\beta{\ln}(1/x)[R_{k,g}^{(2)}(1,\xi_{r})+4C_{A}R_{k,g}^{(1)}(1,\xi_{r})L_{\mu}+8C_{A}^{2}B_{k,g}^{(0)}(1,\xi_{r})L^{2}_{\mu}],
\end{eqnarray}
with
\begin{eqnarray}
R_{2,g}^{(2)}(1,\xi_{r})&=&
\frac{32}{27}C_{A}^{2}[46+(71-92a)J(\xi_{r})+3(13-10\xi_{r})I(\xi_{r})-9(1-\xi_{r})K(\xi_{r})],\nonumber\\
R_{L,g}^{(2)}(1,\xi_{r})&=&\frac{64}{27}C_{A}^{2}x_{2}\{34+240\xi_{r}-[3+136\xi_{r}+480\xi_{r}^{2}]J(\xi_{r})+3[3+4\xi_{r}(1-6\xi_{r})]I(\xi_{r})+18\xi_{r}(1+3\xi_{r})K(\xi_{r})\},\nonumber\\
R_{2,g}^{(1)}(1,\xi_{r})&=&
\frac{8}{9}C_{A}[5+(13-10\xi_{r})J(\xi_{r})+6(1-\xi_{r})I(\xi_{r})],\nonumber\\
R_{L,g}^{(1)}(1,\xi_{r})&=&-\frac{16}{9}C_{A}x_{2}\{1-12\xi_{r}-[3+4\xi_{r}(1-6\xi_{r})]J(\xi_{r})+12\xi_{r}[1+3\xi_{r}]I(\xi_{r})\},\nonumber\\
\end{eqnarray}
where
\begin{eqnarray}
K(\xi_{r})&=&-\sqrt{x_{2}}~[4(\zeta_{3}+\mathrm{Li}_{3}(-t)-\mathrm{Li}_{2}(-t){\ln}t-2S_{1,2}(-t))+2{\ln}(\xi_{r}x_{2})(\zeta_{2}+2\mathrm{Li}_{2}(-t))\nonumber\\
&&-\frac{1}{3}{\ln}^{3}t
-{\ln}^{2}(\xi_{r}x_{2}){\ln}t+{\ln}(\xi_{r}x_{2}){\ln}^{2}t],\nonumber\\
I(\xi_{r})&=&-\sqrt{x_{2}}~
[\zeta_{2}+\frac{1}{2}{\ln}^{2}t-{\ln}(\xi_{r}x_{2}){\ln}t+2\mathrm{Li}_{2}(-t)],\nonumber\\
J(\xi_{r})&=&-\sqrt{x_{2}}~ {\ln}t,\nonumber\\
t&=&\frac{1-\sqrt{x_{2}}}{1+\sqrt{x_{2}}},\nonumber\\
x_{2}&=&\frac{1}{1+4\xi_{r}},\nonumber\\
L_{\mu}&=&\ln{\frac{4m_{f}^{2}}{\mu_{r}^{2}}},\nonumber\\
\end{eqnarray}
where
\begin{eqnarray}
\mathrm{Li}_{2}(x)&=&-\int_{0}^{1}\frac{dy}{y}{\ln}(1-xy),\nonumber\\
\mathrm{Li}_{3}(x)&=&-\int_{0}^{1}\frac{dy}{y}{\ln}(y){\ln}(1-xy),\nonumber\\
S_{1,2}(x)&=&\frac{1}{2}\int_{0}^{1}\frac{dy}{y}{\ln}^{2}(1-xy),\nonumber\\
\end{eqnarray}
are the dilogarithmic function $\mathrm{Li}_{2}(x)$, the
trilogarithmic
function $\mathrm{Li}_{3}(x)$ and Nilsen Polylogarithm $S_{1,2}(x)$.\\


\section{References}

1. V.N.Gribov, B.L.Ioffe, and I.Y.Pomeranchuk,  Sov.J.Nucl.Phys.
{\bf2}, 549 (1966).\\
2. B.L.Ioffe, Phys.Lett.B {\bf30}, 123 (1969).\\
3. J.J.Sakurai, Currents and Mesons, The University of Chicago
Press, 1969.\\
4. D.Schildknecht, Acta Phys.Polon.B {\bf37}, 595 (2006).\\
5. N.N.Nikolaev and B.G.Zakharov, Z.Phys.C {\bf49}, 607 (1991);
Z.Phys.C {\bf53}, 331 (1992).\\
6. C.Ewerz, A.von Manteuffel and O.Nachtmann, J.High Energ.Phys. {\bf03}, 102 (2010).\\
7. D.A.Fagundes and M.V.T.Machado, Phys.Rev.D {\bf107}, 014004
(2023).\\
8. V.P.Goncalves and M.V.T.Machado, Phys.Rev.Lett.{\bf91}, 202002 (2003).\\
9. E.Iancu, A.Leonidov and L.McLerran, Nucl.Phys.A {\bf692}, 583
(2001); Phys.Lett.B {\bf510}, 133 (2001).\\
10. E.Iancu,K.Itakura and S.Munier, Phys.Lett.B {\bf590}, 199
(2004).\\
11. K.Kutak and A.M.Stasto, Eur.Phys.J.C {\bf41}, 343 (2005).\\
12. N.N.Nikolaev and B.G.Zakharov, Phys.Lett.B {\bf332}, 184
(1994);
 N. N. Nikolaev and W. Sch$\ddot{a}$fer, Phys. Rev. D {\bf74}, 014023 (2006).\\
13. V.S.Fadin, E.A.Kuraev and L.N.Lipatov, Phys.Lett.B
\textbf{60}, 50(1975); L.N.Lipatov, Sov.J.Nucl.Phys. \textbf{23},
338(1976); I.I.Balitsky and L.N.Lipatov, Sov.J.Nucl.Phys.
\textbf{28}, 822(1978).\\
14. A.H.Mueller and B.Patel, Nucl.Phys.B {\bf425}, 471 (1994).\\
15. R.S.Thorne, Phys.Rev.D 71, 054024 (2005); M.A.Betemps and M.V.T.Machado, Eur.Phys.J.C {\bf65}, 427 (2010).\\
16. R. D. Ball and S. Forte, Phys. Lett. B 335, 77 (1994).\\
17. A.D.Bolognino, A.Szczurek and W. Sch$\ddot{a}$fer, Phys.Rev.D
{\bf101},  054041 (2020); A.D.Bolognino, F.G.Celiberto,
D.Y.Ivanov, A. Papa, W. Sch$\ddot{a}$fer and A. Szczurek, Eur.
Phys.J.C {\bf81}, 846 (2021); A.D.Bolognino, F.G.Celiberto,
D.Y.Ivanov and A. Papa,  Eur. Phys.J.C {\bf78}, 1023 (2018);
A.D.Bolognino, F.G.Celiberto, M.Fucilla, Dmitry Yu. Ivanov,
A.Papa, W.Schafer and A.Szczurek, International Conference on Hadron Spectroscopy and Structure in memoriam Simon Eidelman (HADRON2021), (2021).\\
18. I.P.Ivanov and N.N.Nikolaev, Phys.Rev.D {\bf65}, 054004
(2002).\\
19. M.A.Kimber, J.Kwiecinski, A.D.Martin and A.M.Stasto,
Phys.Rev.D
{\bf62}, 094006 (2000).\\
20. T.Goda, K.Kutak and S.Sapeta, Nucl.Phys.B {\bf990}, 116155 (2023).\\
21. M.A.Kimber, A.D.Martin and M.G.Ryskin, Eur.Phys.J.C {\bf12},
655 (2000).\\
22. B.W.Xiao, F.Yuan and J.Zhou, Nucl.Phys.B {\bf921}, 104
(2017).\\
23. K. Golec-Biernat and S.Sapeta, J.High Energ. Phys. {\bf03},
102 (2018).\\
24. LHeC Collaboration and FCC-he Study Group, P. Agostini et al.,
J. Phys. G: Nucl. Part. Phys. {\bf48}, 110501(2021).\\
25.  E. Laenen et al., Phys. Lett. B{\bf291}, 325 (1992); E.
Laenen et al., Nucl. Phys. B{\bf392}, 162 (1993); S. Riemersma, J.
Smith,W.L. van Neerven, Phys. Lett. B{\bf347}, 143 (1995); S.
Alekhin et al., Phys. Rev.D{\bf81}, 014032 (2010); S. Alekhin and
S. Moch, Proc. of DIS2011, (2011); S. Alekhin, J.
Bl$\ddot{\mathrm{u}}$mlein, S. Moch, Phys. Rev. D{\bf86}, 054009
(2012); S. Alekhin et al., Phys. Rev. D{\bf96}, 014011 (2017); S.
Alekhin, J. Bl$\ddot{u}$mlein, S. Klein and  S. Moch, Proc. of
DIS2009, (2009); M. Gl$\ddot{\mathrm{u}}$ck et al., Phys. Lett.
B{\bf664}, 133 (2008); H.L. Lai et al., Phys. Rev. D{\bf82},
074024 (2010); A.D. Martin et al., Eur. Phys. J. C{\bf70}, 51
(2010); S. Alekhin and S. Moch,
Phys. Lett. B{\bf699}, 345 (2011).\\
26. S. Forte et al., Nucl. Phys. B{\bf834}, 116 (2010); R.D. Ball
et al. [NNPDF Collaboration], Nucl. Phys. B{\bf849}, 296 (2011);
R.D. Ball et al. [NNPDF Collaboration], Nucl. Phys. B{\bf855}, 153
(2012); R.D.
Ball et al. Eur.Phys.J.C {\bf78}, 321 (2018).\\
27. R.Thorne, Phys.Rev.D{\bf73}, 054019 (2006); R.Thorne,
Phys.Rev.D{\bf86}, 074017 (2012).\\
28. R.S.Thorne, DIS1998, (1998); A.D.Martin W.J.Stirling and R.S.Thorne, Phys.Lett.B {\bf 636}, 259(2006).\\
29. J.Bl$\ddot{\mathrm{u}}$mlein, A.De Freitas, C.Schneider and
K.Sch$\ddot{\mathrm{o}}$nwald, Phys. Lett.B {\bf782},
 362(2018).\\
30.  S.Alekhin, J. Bl$\ddot{\mathrm{u}}$mlein and S. Moch, Phys. Rev. D {\bf102}, 054014 (2020).\\
31. A.V.Kotikov and G.Parente, Nucl.Phys.B {\bf549}, 242 (1999).\\
32. A.Yu.Illarionov, A.V.Kotikov and G.Parente, Phys.Part.Nucl.
{\bf39}, 307 (2008).\\
33. L.Mankiewicz, A.Saalfeld and T.Weigl, Phys.Lett.B {\bf393},
175 (1997).\\
34. A.V.Kotikov, A.V.Lipatov and P.Zhang, Phys.Rev.D {\bf104}, 054042 (2021).\\
35. G.R.Boroun and B.Rezaei, arXiv[hep-ph]:2309.04832 (will be appear in EPJA).\\
36. G.Beuf, T.Lappi and R.Paatelainen, Phys.Rev.D {\bf104}, 056032
(2021).\\
37. G.Beuf, Phys.Rev.D {\bf85}, 034039 (2012).\\
38. C.Ewerz, A.von Manteuffel and O.Nachtmann, Phys.Rev.D {\bf77},
074022 (2008); M.Niedziela and
M.Praszalowicz, Acta Physica Polonica B{\bf46}, 2018 (2015).\\
39. C.Ewerz, A.von Manteuffel, O.Nachtmann and A.Schoning, Phys.
Lett.B {\bf720}, 181 (2013); C.Ewerz and O.Nachtmann, Phys.Lett.B {\bf648}, 279 (2007).\\
40. B.Rezaei and G.R.Boroun, Phys.Rev.C {\bf101}, 045202 (2020); G.R.Boroun and B.Rezaei,
Phys.Rev.C {\bf103}, 065202 (2021); Phys.Letts.B {\bf816}, 136274 (2021).\\
41. G.R.Boroun,  Eur.Phys.J.A {\bf57}, 219 (2021).\\
42. M.Klein, arXiv:1802.04317; M.Klein, Ann.Phys.{\bf528},
138(2016).\\
43. G.R.Boroun, M.Kuroda and D.Schildknecht, arXiv: 2206.05672.\\
44.  M.Kuroda and D.Schildknecht, Phys.Rev. D {\bf96}, 094013
(2017); Phys.Rev. D {\bf85}, 094001 (2012); Int. J. Mod. Phys. A
{\bf 31}, 1650157 (2016).\\
45. LHC Higgs Cross Section Working Group collaboration,
CERN Yellow Reports: Monographs Volume 2/2017 (CERN--2017--002-M).\\
46. NNPDF Collaboration (Ball R. D. et al.), Eur.Phys.J.C {\bf77},
663 (2017).\\
47. S.Zarrin and S.Dadfar, Phys.Rev.D {\bf106}, 094007 (2022).\\
48. J. Lan et al., Phys. Rev. D {\bf102}, 014020 (2020).\\
49. H. Khanpour, Nucl.Phys.B {\bf958}, 115141 (2020).\\
50. G.R.Boroun, Chinese Physics C {\bf45}, 063105 (2021);
Eur.Phys.J.Plus {\bf138}, 252 (2023); Phys.Letts.B {\bf838}, 137712 (2023).\\
51. F.D.Aaron et al. (H1 Collaboration), Eur.Phys.J.C {\bf65}, 89
(2010).\\
52. N.N.Nikolaev and V.R.Zoller, Phys.Atom.Nucl {\bf73}, 672
(2010); N.N.Nikolaev, J.Speth and V.R.Zoller, Phys.Lett.B
{\bf473}, 157 (2000);  R.Fiore, N.N.Nikolaev and V.R.Zoller, JETP
Lett {\bf90}, 319 (2009); A.Y.Illarionov and A.V.Kotikov,
Phys.Atom.Nucl. {\bf75}, 1234 (2012) ; N.Ya.Ivanov and B.A.Kniehl,
Eur.Phys.J.C {\bf59}, 647 (2009); N.Ya.Ivanov, Nucl.Phys.B
{\bf814}, 142(2009); J.Blumlein et al., Nucl.Phys.B {\bf755}, 272
(2006); A.V.Kotikov, arXiv[hep-ph]: 1212.3733; G.R.Boroun and
B.Rezaei, Int.J.Mod.Phys.E {\bf24}, 1550063 (2015); G.R.Boroun and B.Rezaei, Nucl.Phys.A {\bf929}, 119 (2014); G.R.Boroun, Eur. Phys. J. Plus {\bf137}, 1212 (2022).\\
53. N.A.Abdulov, A.V.Kotikov and A.V.Lipatov, JETP Lett. {\bf117},
401 (2023).\\
54. C.Adloff et al. (H1 Collaboration), Eur.Phys.J.C {\bf21}, 33
(2001).\\
55. G.R.Boroun and B.Rezaei, EPL {\bf133}, 61002 (2021).\\
56. C. Adloff et al. (H1 Collaboration), Z.Phys.C {\bf72}, 593
(1996).\\

\end{document}